\documentclass[12pt]{article}
\usepackage{amsfonts}
\usepackage{amssymb}

\def\hybrid{\topmargin -20pt    \oddsidemargin 0pt
        \headheight 0pt \headsep 0pt
        \textwidth 6.35in       
       \textheight 9.25in       
        \marginparwidth .875in
        \parskip 5pt plus 1pt   \jot = 1.5ex}

\hybrid

\def\baselinestretch{1.2}

\catcode`\@=11

\def\marginnote#1{}
\newcount\hour
\newcount\minute
\newtoks\amorpm
\hour=\time\divide\hour by60
\minute=\time{\multiply\hour by60 \global\advance\minute by-\hour}
\edef\standardtime{{\ifnum\hour<12 \global\amorpm={am}%
        \else\global\amorpm={pm}\advance\hour by-12 \fi
        \ifnum\hour=0 \hour=12 \fi
        \number\hour:\ifnum\minute<10 0\fi\number\minute\the\amorpm}}
\edef\militarytime{\number\hour:\ifnum\minute<10 0\fi\number\minute}

\def\draftlabel#1{{\@bsphack\if@filesw {\let\thepage\relax
   \xdef\@gtempa{\write\@auxout{\string
      \newlabel{#1}{{\@currentlabel}{\thepage}}}}}\@gtempa
   \if@nobreak \ifvmode\nobreak\fi\fi\fi\@esphack}
        \gdef\@eqnlabel{#1}}
\def\@eqnlabel{}
\def\@vacuum{}
\def\draftmarginnote#1{\marginpar{\raggedright\scriptsize\tt#1}}

\def\draft{\oddsidemargin -.5truein
        \def\@oddfoot{\sl preliminary draft \hfil
        \rm\thepage\hfil\sl\today\quad\militarytime}
        \let\@evenfoot\@oddfoot \overfullrule 3pt
        \let\label=\draftlabel
        \let\marginnote=\draftmarginnote
   \def\@eqnnum{(\theequation)\rlap{\kern\marginparsep\tt\@eqnlabel}%
\global\let\@eqnlabel\@vacuum}  }

\def\preprint{\twocolumn\sloppy\flushbottom\parindent 2em
        \leftmargini 2em\leftmarginv .5em\leftmarginvi .5em
        \oddsidemargin -.5in    \evensidemargin -.5in
        \columnsep .4in \footheight 0pt
        \textwidth 10.in        \topmargin  -.4in
        \headheight 12pt \topskip .4in
        \textheight 6.9in \footskip 0pt
        \def\@oddhead{\thepage\hfil\addtocounter{page}{1}\thepage}
        \let\@evenhead\@oddhead \def\@oddfoot{} \def\@evenfoot{} }

\def\numberbysection{\@addtoreset{equation}{section}
        \def\theequation{\thesection.\arabic{equation}}}

\def\underline#1{\relax\ifmmode\@@underline#1\else
        $\@@underline{\hbox{#1}}$\relax\fi}

\def\titlepage{\@restonecolfalse\if@twocolumn\@restonecoltrue\onecolumn
     \else \newpage \fi \thispagestyle{empty}\c@page\z@
        \def\thefootnote{\fnsymbol{footnote}} }

\def\endtitlepage{\if@restonecol\twocolumn \else \newpage \fi
        \def\thefootnote{\arabic{footnote}}
        \setcounter{footnote}{0}}  

\catcode`@=12
\relax

\def\figcap{\section*{Figure Captions\markboth
        {FIGURECAPTIONS}{FIGURECAPTIONS}}\list
        {Figure \arabic{enumi}:\hfill}{\settowidth\labelwidth{Figure
999:}
        \leftmargin\labelwidth
        \advance\leftmargin\labelsep\usecounter{enumi}}}
 \relax
\def\tablecap{\section*{Table Captions\markboth
        {TABLECAPTIONS}{TABLECAPTIONS}}\list
        {Table \arabic{enumi}:\hfill}{\settowidth\labelwidth{Table
999:}
        \leftmargin\labelwidth
        \advance\leftmargin\labelsep\usecounter{enumi}}}
 \relax
\def\reflist{\section*{References\markboth
        {REFLIST}{REFLIST}}\list
        {[\arabic{enumi}]\hfill}{\settowidth\labelwidth{[999]}
        \leftmargin\labelwidth
        \advance\leftmargin\labelsep\usecounter{enumi}}}
 \relax
\makeatletter
\newcounter{pubctr}
\def\publist{\@ifnextchar[{\@publist}{\@@publist}}
\def\@publist[#1]{\list
        {[\arabic{pubctr}]\hfill}{\settowidth\labelwidth{[999]}
        \leftmargin\labelwidth
        \advance\leftmargin\labelsep
        \@nmbrlisttrue\def\@listctr{pubctr}
        \setcounter{pubctr}{#1}\addtocounter{pubctr}{-1}}}
\def\@@publist{\list
        {[\arabic{pubctr}]\hfill}{\settowidth\labelwidth{[999]}
        \leftmargin\labelwidth
        \advance\leftmargin\labelsep
        \@nmbrlisttrue\def\@listctr{pubctr}}}
 \relax
\makeatother
\newskip\humongous \humongous=0pt plus 1000pt minus 1000pt

\newif\ifdtup

\relax

\def\be{\begin{equation}}
\def\ee{\end{equation}}
\def\ba{\begin{eqnarray}}
\def\ea{\end{eqnarray}}

\def\a{\alpha}

\def\g{\gamma}

\def\d{\delta}

\def\e{\epsilon}

\def\th{\theta}

\def\m{\mu}
\def\n{\nu}

\def\Om{\Omega}

\def\qq{\qquad}

\def\IR{\relax{\rm I\kern-.18em R}}

\def \ha {{1\over 2}}

\def \ov {\over}

\def\diag{{\rm diag}}

\def\IR{\relax{\rm I\kern-.18em R}}
\def\inv{^{\raise.15ex\hbox{${\scriptscriptstyle -}$}\kern-.05em 1}}

\begin{document}

\renewcommand{\theequation}{\arabic{equation}}

\newcommand{\beq}{\begin{equation}}
\newcommand{\eeq}[1]{\label{#1}\end{equation}}
\newcommand{\ber}{\begin{eqnarray}}
\newcommand{\eer}[1]{\label{#1}\end{eqnarray}}
\newcommand{\eqn}[1]{(\ref{#1})}
\begin{titlepage}
\begin{center}

\hfill CALT-68-2418\\
\vskip -.1 cm
\hfill hep--th/0212056\\

\vskip .5in

{\large \bf PP-waves from rotating and continuously distributed D3-branes}
\vskip 0.4in

{\bf Andreas Brandhuber}${}^1$
and
{\bf Konstadinos Sfetsos}${}^2$
\vskip 0.1in
{\em ${}^1\!$Department of Theoretical Physics\\
     California Institute of Technology\\
     Pasadena, CA 91125, USA\\
\footnotesize{\tt andreas@theory.caltech.edu}}\\
\vskip .2in
{\em ${}^2\!$Department of Engineering Sciences, University of Patras\\
26110 Patras, Greece\\
\footnotesize{\tt sfetsos@mail.cern.ch, des.upatras.gr}}\\

\end{center}

\vskip .3in

\centerline{\bf Abstract}

\noindent
We study families of PP-wave solutions of type-IIB supergravity that have
(light-cone) time dependent metrics and RR five-form fluxes. They arise as
Penrose limits of supergravity solutions that correspond
to rotating or continuous distributions of D3-branes. In general,
the solutions preserve sixteen supersymmetries.
On the dual field theory side these backgrounds describe the BMN limit of
$\mathcal{N}=4$ SYM when some scalars in the field theory have non-vanishing
expectation values. We study the perturbative string spectrum
and in several cases we are able to determine it exactly for the bosons as
well as for the fermions. We find that there are special states for
particular values of the light-cone constant $P_+$.

\vskip .5in
\noindent
November 2002\\
\end{titlepage}
\vfill
\eject

\def\baselinestretch{1.2}
\baselineskip 20 pt
\noindent

\section{Introduction}

Recently, an interesting testing ground for the AdS/CFT correspondence was
unraveled \cite{bmn} which allows to make precise predictions and comparisons
on both sides of the duality.
This is based on the observation that string
theory is exactly solvable in particular PP-wave backgrounds
\cite{metsaev,mettse}.
Furthermore, these backgrounds are Penrose
limits \cite{Penrose} of
the $AdS_d \times S^k$ space-times, that where
intensively studied in the context of the AdS/CFT correspondence. This
geometric limit translates to a truncation of the gauge theory to operators
of large $\mathcal{R}$ charge. The aim of this paper, with these new insights
in mind, is to investigate
supergravity backgrounds that have already been used to test the AdS/CFT
correspondence successfully. These backgrounds describe deformations of
the $\mathcal{N} \! = \! 4$ supersymmetric Yang--Mills theory by turning on
expectation values of scalar fields.
The supergravity solutions involve non-trivial
RR five-form fluxes and the metric, and they preserve 16 supersymmetries.
In the Penrose limit the metric and the five-form depend in general on the
light-cone time, unlike the PP-waves with higher supersymmetry, and
no supersymmetry enhancement is observed. In addition, we will study
backgrounds that correspond to the same Yang--Mills theory at finite
temperature and strong 't Hooft coupling which break supersymmetry.
In these cases the Penrose limit restores 16 of the broken supersymmetries.

Besides flat space-time, type-IIB supergravity \cite{schwarz}
admits two extra maximally supersymmetric solutions, namely
$AdS_5 \times S^5$ \cite{compp} and the PP-wave \cite{Blau2}, both
with the RR five-form flux turned on.
The latter solution is the Penrose limit of the
former one \cite{Blaupen}\footnote{
Penrose limits in order to obtain PP-wave solutions in string theory were
first performed in \cite{Sfe},
in relation to WZW and gauged WZW models based on
non-semisimple groups (initiated in \cite{nappi}),
and also in \cite{Gueven} for the
field equations of the low energy perturbative string theories.}
and as shown in \cite{bmn} string theory in this background is dual
to a corner of the $\mathcal{N} \! = \! 4$ supersymmetric Yang--Mills theory
just as the Penrose limit blows up an infinitesimal region around a
null-geodesic in the full  $AdS_5 \times S^5$ geometry.
Naturally, one is led to the question if
this can be generalized to theories with less supersymmetry and/or broken
conformal symmetry. Penrose limits of geometries describing RG fixed points
have been studied recently which led to PP-waves with various amounts of
supersymmetry ranging from 16 to 32, where in many cases
the supersymmetry compared to the original background is enhanced.
In all these cases the metric
and the fluxes are static (at least after a simple coordinate transformation)
and string theory can be solved exactly \cite{cvetic}-\cite{casero}.
Harder and more interesting is the case of
Penrose limits of RG flows which generically lead to PP-waves and fluxes that
depend on the light-cone time. These waves always preserve 16 supercharges
and generically string theory cannot be solved because the world-sheet
theory is interacting. The aim of this paper is to study the Penrose limits of
one of the simplest
RG flows that can be studied using the AdS/CFT correspondence and
investigate aspects of string propagation in the corresponding  PP-waves.
It turns out that even for these simple models
the interactions are quite complicated but in several cases at least the
spectrum of string excitations can be worked out explicitly.

This paper is organized as follows:
In section 2 we will present the general features of the PP-waves of the form
we will
discuss later and count the number of supersymmetries they preserve.
In section 3 we will take two different Penrose limits on type-IIB backgrounds
representing near-extremal rotating D3-branes, or continous distributions of
D3-branes and we will describe the
resulting PP-waves in detail. Section 4 is devoted to an analysis of the
bosonic and fermionic modes of the Green--Schwarz string in these PP-waves.
We exhibit cases for which the perturbative string spectra can be
found exactly, despite the non-trivial dependence of our backgrounds on
the light-cone time.
Finally, in section 5 we study D3-brane distributions in a limit where the
size of the distribution goes to infinity, as well as their Penrose limit.
We end the paper with concluding remarks in section 6.

\section{A class of PP-waves and supersymmetry}

Before going into explicit examples let us summarize some basic features of
the class of PP-wave solutions that we are going to construct.
First of all the dilaton and axion
are constant and the Ramond and Neveu--Schwarz two-form gauge fields vanish.
The general form of the remaining, non-trivial fields is
\ba\nonumber
&& ds^2 = 2 du dv +
\sum_{i,j=1}^8 F_{ij}(u) x^i x^j du^2 + \sum_{i=1}^8 dx^i dx^i ~,\\
&& \mathcal{F}_5 = f(u) du \wedge (1 + \ast_8)(dx_1 \wedge dx_2 \wedge dx_3
\wedge dx_4) ~,
\label{general}
\ea
where $\ast_8$ denotes the Hodge star operator acting on the eight directions
transverse to $u,v$.
One can easily read off the isometries from \eqn{general}.
First there is always a shift symmetry in $v$, whereas a shift of $u$
is a symmetry only if $F_{ij}$ and $f$ are constant.
In addition,
depending on the $F_{ij}$, there can be symmetries rotating the $x_i$.
However,
the form of the Ramond five-form $\mathcal{F}_5$ restricts the maximal
symmetry to $SO(4) \times SO(4)$ if $F_{11}=F_{22}=F_{33}=F_{44}$,
$F_{55}=F_{66}=F_{77}=F_{88}$ and all other $F_{ij}=0$.
For less symmetric configurations the symmetry is reduced accordingly.
We will not discuss any further, but simply note here,
that backgrounds where the
space transverse to the light-cone directions
is replaced by a curved manifold and
the $F_{ij}$ depend only on these transverse coordinates
have been studied in \cite{floratos}-\cite{Kim}.

The metric in \eqn{general} can be written as
\be\label{vbmetric}
ds^2 = 2 e^+ e^- + \sum_{i=1}^8 e^i e^i ~,
\ee
by introducing the Zehnbeine $e^a = e^a_M dx^M$:
\be\label{zehnbein}
e^+ = du ~,\qq e^- = dv + \ha F_{ij} x^i x^j du ~,\qq  e^i = dx^i \ .
\ee
The non-vanishing components of the spin connection $\omega_{ab}$ are
\be
\label{spinconn}
\omega_{+i} = \omega^{-i} = F_{ij} x^j du \ .
\ee
The only non-vanishing components of the Riemann and Ricci tensors are
\be
R_{+i+j}=-F_{ij} \ , \qq R_{++} = {-F_i}^i \ ,
\ee
simplifying tremendously the classical type-IIB field equations to
\be\label{einstein}
{-F_i}^i = 8 f(u)^2 ~.
\ee

We can determine the number of supersymmetries preserved
by the PP-wave solution \eqn{general} (for $F_{ij}$ and $f$ constants,
this was done in \cite{Blau2}).
For this purpose we have to set to zero the type-IIB supergravitity
variations \cite{schwarz}
of the dilatino, $\lambda$, and the gravitino, $\psi_M$.
For our particular backgrounds the complex three-form vanishes
identically, and, therefore, the dilatino equation is trivially solved
whereas the gravitino equation reduces to
\be\label{gravitino}
\delta \psi_M = D_M \epsilon + \frac{i}{480}
\mathcal{F}_{PQRST} \Gamma^{PQRST} \Gamma_M \epsilon ~,
\ee
where
$D_M \epsilon = \partial_M \epsilon + \frac{1}{4} \omega^{PQ}_M
\Gamma^{PQ} \epsilon$
and we set $\gamma^{11} \psi_M = \psi_M$, so that
$\gamma^{11} \epsilon = \epsilon$.

The covariant derivatives appearing in \eqn{gravitino} become
\be\label{covderiv}
D_- = \partial_- ~,\qq D_i = \partial_i ~,\qq
D_+ = \partial_+ + \ha F_{ij} x^j \g_- \g_i  ~,
\ee
where the $\gamma^a$ are ten-dimensional flat space Gamma matrices
(for a convenient basis, see, for instance, \cite{mettse})
and where we have defined
\be\label{gammapm}
\gamma^\pm = (\pm \gamma^0 + \gamma^9)/\sqrt{2} ~.
\ee
The last term in the gravitino equation \eqn{gravitino} using \eqn{general}
is
\be
\frac{i}{480} \mathcal{F}_{PQRST} \Gamma^{PQRST}
\gamma_M = \frac{i}{4} f(u) (P_1+P_2) \gamma_- \gamma_M ~,
\ee
where
\be
P_1= \g_1 \g_2 \g_3 \g_4\ , \qq P_2= \g_5 \g_6 \g_7 \g_8\ ,
\label{porp}
\ee
are products of Gamma matrices acting as projection
operators in the four-dimensional
subspaces defined by the five-form ansatz in \eqn{general}. The obey
\ba
&& P_1^2 = P_2^2 =1\ , \qq [P_1,P_2]=[P_{1,2},\g_{\pm}]= 0\ ,
\nonumber\\
&&\{P_1,\g_i\}= [P_1,\g_{i+4}]=0 \ ,\qq i=1,2,3,4\ ,
\label{prpp}\\
&&\{P_2,\g_i\}= [P_2,\g_{i-4}]=0 \ ,\qq i=5,6,7,8\ .
\nonumber
\ea
Combined with \eqn{covderiv}, we find \eqn{gravitino} in components
\ba
\partial_- \epsilon & = & 0 ~,\nonumber \\
\partial_i \epsilon & = & - \frac{i}{4} f(u) \Omega^i \epsilon ~,
\label{spineq} \\
\partial_+ \epsilon & = & -\frac{1}{2}  F_{ij} x^j \gamma_- \g_i
\epsilon - \frac{i}{4} f(u) (P_1+P_2) \gamma_- \g_+ \epsilon ~ ,
\nonumber
\ea
where
\be
\Om_i= (P_1+P_2) \g_- \g_i \ ,
\ee
obeying
\be
[P_1+P_2,\Om_i] = 2 (1+P_1 P_2) \g_- \g_i\ .
\ee
The first condition in \eqn{spineq} states that $\epsilon$ is a function
of $x^i$ and $x^+=u$ only. The second condition can be solved by
using the fact that $\Omega^i \Omega^j = 0$, which implies that
$\partial_{i}\partial_j \epsilon = 0$,
{\it i.e.} $\epsilon$ is linear in the $x^i$.
The solution is
\be\label{firstep}
\epsilon = (1 - \frac{i}{4} f(u) x^j \Omega^j) \chi(u) ~,
\ee
which we insert in the third condition in \eqn{spineq} resulting in
a linear equation in $x^i$. Hence we obtain two condition that read
\be\label{zero}
\left(\partial_u + \frac{i}{4} f(u) (P_1 + P_2) \gamma_- \g_+\right) \chi = 0
\ ,
\ee
and
\be\label{linear}
x^i \left(i f^\prime(u) (P_1+P_2)\g_i -2 F_{ij} \g_j -2 f^2 \g_i\right)
\g_- \chi =0\ .
\ee
In the various manipulations we have used repeatedly that $\g_\pm^2=0$ and
that
\be
(1\mp P_1P_2)\chi = \g_{\pm}\g_{\mp}\chi\ ,
\label{pp29}
\ee
which is a consequence of the chirality condition
$\gamma^{11} \e = \e$ and the
fact that $\gamma^{11}$ anti-commutes with all Gamma matrices.

For generic $f$ and $F_{ij}$ this equation has 16 solutions
given by  $\gamma^+ \chi = \gamma_- \chi = 0$ and the explicit
$u$-dependence is easily found by integrating \eqn{zero}.
The maximal supersymmetric solution with 32 supercharges corresponds to
constant $f$ and $F_{ij}$ related by $F_{ij}=-f^2 \d_{ij}$, $i=1,\ldots,8$.
In cases where the function $f(u)$ is not a constant
there exist in general no additional real supersymmetric solutions.
To see this, note that for spinors obeying $\g_-\chi=0$ we have,
due to \eqn{pp29}, one of
two possibilities: $P_1\chi=P_2\chi=\pm \chi$. Therefore,
any additional supersymmetries must meet one of the two
remaining possibilities:
$P_1\chi=-P_2\chi=\pm \chi$. However, using the fact that $\g_i$ commutes
with one of the projectors among $P_1$, $P_2$ and anti-commutes with the other
\eqn{prpp}, we see that
the eq. \eqn{linear} will give rise to complex $F_{ij}$ and, hence,
to complex metrics which are physically unacceptable.

\section{Rotating D3-branes and Penrose limits}

In this section we construct Penrose limits of the supergravity solutions of
rotating D3-branes \cite{rott,russosfe}.
The most general solution is characterized by five
constants:
the number of D3-branes, $N$, three rotation parameters, $r_{1,2,3}$, and
the near-extremality parameter, $\mu$, which is related to the Hawking
temperature
of the black brane solution. For general $r_i$ the isometry of the
transverse space,
which is related to the $\mathcal{R}$-symmetry in the dual field theory,
is reduced
$SO(6) \to SO(2)\times SO(2)\times SO(2)$.
We will not consider the most general case, but we will
restrict our attention to backgrounds with
only one non-vanishing rotation parameter: $r_1 = r_0,~ r_2 = r_3 = 0$.
Then the symmetry of the solution is $ISO(3,1)\times SO(4)\times SO(2)$.
The explicit form of the metric and RR five-form field strength of
the supergravity solution,
in the field theory limit, is \cite{russosfe}
\ba
ds^2 &=& H^{-1/2}\left( -f dt^2 + dx_1^2 + dx_2^2 +dx_3^2\right)
+ H^{1/2} \Bigg({dr^2\over f_1}  + (r^2+r_0^2 \cos^2\th)\ d\th^2
\nonumber\\
&& + (r^2+r_0^2) \sin^2\th\ d\phi^2 +r^2 \cos^2\th\ d\Omega_3^2
-{2 \m^2 r_0\over R^2} \sin^2\th d t d\phi\Bigg) ~, \nonumber \\
\mathcal{F}_5 & = & d C_4 + \ast d C_4  ~~  \mathrm{with}
\label{r2uu1} \\
C_4  & = &  (H^{-1}  dt + r_0
\m^2/R^2 \sin^2\th d\phi) \wedge dx^1 \wedge dx^2 \wedge dx^3 ~,
\nonumber
\ea
and
\ba
H & =&  {R^4 \over r^2(r^2+r_0^2\cos^2\th)}\ ,
\nonumber \\
 f & =&  1-{\m^4\ov r^2(r^2+r_0^2\cos^2\th)}\ ,
\label{defff}\\
f_1& =&  {r^4+r_0^2 r^2 -\m^4\ov
r^2(r^2+r_0^2\cos^2\th)}\ ,
\nonumber
\ea
where $R^4 = 4 \pi g_s N$
and $r_0$ is the angular momentum parameter.
Note that we set $\alpha'=1$ throughout the paper.
The location of the horizon is given by the positive
root of the equation $r^4+r_0^2 r^2 -\m^4=0$
\be
r_H^2 = \ha \left( \sqrt{r_0^4 + 4 \m^4}-r_0^2\right)\ ,
\label{jsab}
\ee
and the Hawking temperature associated with \eqn{r2uu1} is
\be
T_H = {r_H\ov 2\pi R^2 \m^2} \sqrt{r_0^4+4 \m^4}\ .
\label{sh21}
\ee
The background \eqn{r2uu1} has been used in various studies within the AdS/CFT
correspondence in \cite{Csaki,russosfe,brandhusfe}.

In order to take a Penrose limit we have to choose a null geodesic on
this space.
This will in general
involve the time directions $t$, the radius $r$ and some directions on
the transverse
$S^5$. In contrast to the maximally supersymmetric case the $S^5$ is
now squashed,
hence the isometry is reduced, and there are several inequivalent
choices for the
geodesics.
To make life a bit easier we only consider geodesics with constant $\theta$
and this leaves two consistent possibilities: $\th = \pi/2$ or $\th=0$.

\subsection{Null geodesics}

The null geodesics that are relevant for taking Penrose limits
involve the directions $t$, $r$ and a particular angular direction
which we denote for the moment by $\alpha$. All other angular
directions and the flat spatial coordinates along the brane $x_{1,2,3}$
are taken to be constant.
After a convenient rescaling $x^\m\to R^2 x^\m$,
we find a three-dimensional effective metric
\be\label{flox}
ds_3^2/R^2 = -G_t dt^2 + G_r dr^2 + G_\alpha d\alpha^2 -
2 G_{t\alpha} dt d\alpha\ ,
\ee
which contains an off-diagonal term due to the angular momentum.

We are looking for null geodesics parametrized by $r,t,\alpha$ as a
function of the proper time $\tau$.
Conservation of energy and angular momentum implies that
\be
G_t \dot t + G_{t\alpha} \dot \alpha = E \equiv 1\ ,
\qq G_{\alpha} \dot \alpha - G_{t\alpha}
\dot t = J \ ,
\ee
with solution
\be
\dot t = {G_\alpha - J G_{t\alpha} \ov G_{t\alpha}^2 + G_t G_\alpha} \ , \qq
\dot \alpha = {G_{t\alpha} + J G_{t} \ov G_{t\alpha}^2 + G_t G_\alpha} \ .
\ee

These expressions can now be fed back into the effective line element
\eqn{flox} and
requiring the geodesic to be null yields a differential equation for $r$
\be\label{geod}
\dot r^2 = {G_\alpha -2 J G_{t\alpha} -J^2 G_t\ov
G_r (G_{t\alpha}^2 + G_t G_\alpha)}  \ .
\ee

For completeness we present here also the general form of
a particular change of coordinates that is important for taking the Penrose
limit
(see also \cite{papa})
\ba
&& dr =\sqrt{G_\alpha -2 J G_{t\alpha} -J^2 G_t\ov
G_r (G_{t\alpha}^2 + G_t G_\alpha)} du \ ,
\nonumber\\
&& dt = {G_{\alpha} - J G_{t\alpha}\ov G_{t\alpha}^2 + G_t G_\alpha}du
- \frac{1}{R^2} dv + \frac{J}{R} dx \ ,
\label{chava3} \\
&& d\alpha ={G_{t\alpha} + J G_{t}\ov G_{t\alpha}^2 + G_t G_\alpha} du +
\frac{1}{R} dx \ .
\nonumber
\ea
Comparing with \eqn{geod} we see that $u$ plays the role of the proper time in
the null geodesic.

\subsection{Geodesics at $\th=\pi/2$}

When $\th=\pi/2$ we see that the $S^3$ part in \eqn{r2uu1} vanishes and
the non-trivial angular direction is $\alpha \equiv \phi$. In this case
the metric elements of the three-dimensional effective metric \eqn{flox}
are
\be
G_t = r^2 - {\m^4\ov r^2} \ ,
\quad G_r = {r^2 \ov r^4 +r_0^2 r^2 -\m^4}\ ,
\quad G_\phi = 1+{r_0^2\ov r^2}\ ,\quad G_{t\phi}= {\m^2 r_0\ov r^2} \ .
\ee
The general (real) solution of \eqn{geod} turns out to be
\be\label{fuga}
r^2(u) = \frac{1}{2 J^2}(1 - a \cos 2 J u)\ ,\qq a \equiv
\sqrt{1+4(r_0 - J \mu^2)^2 J^2} \ .
\ee

For the Penrose limit we perform the change of
variables \eqn{chava3}, set
\be
\th = {\pi\ov 2} - {z\ov R} \ ,\qq  x_{1,2,3} \to x_{1,2,3}/R\ ,
\ee
and define $dz^2 + z^2 d\Omega_3^2 = d\vec{x}^2_4$, so that
$z^2 = \vec{x}^2_4$.
Furthermore, we combine the spacelike brane
directions into the three-vector $\vec{x}_3$.
The Penrose limit of \eqn{r2uu1} is then obtained by employing all these
coordinate transformations and taking the limit $R \to \infty$.
The resulting metric is
\be
ds^2 = 2 du dv  + d\vec x_4^2 + A_3(u) d\vec x_3^2 + A_x (u) dx^2
- J^2 \vec x_4^2 du^2 \ ,
\label{penr11}
\ee
where the various functions are
\ba
 && A_x = 1+r_0^2/r^2 - J^2 r^2 = \ha {a^2 \sin^2 2 J u\ov 1-a \cos 2 J u}
\ , \nonumber\\
&& A_3 = r^2 =  {1\ov 2 J^2} (1-a \cos 2 J u )\ ,
\label{penr12}
\ea
 Furthermore, it will be useful to present the metric in a different
form using Brinkman coordinates.\footnote{For a metric of the
form $ds^2 = 2 du dv + A dx^2$ this means that $u\to u$, $v\to v+
{A'\ov 4A} x^2$, $x\to x/\sqrt{A}$, which then gives the metric
$ds^2 =2 dudv + dx^2+ F x^2 du^2 $, where $F= {1 \over 4} A'^2/A^2 +
\ha (A'/A)'$. In the
case of several transverse directions this transformation is trivially
iterated.
If there is a $du^2$ component in the original metric then this is
absorbed in
the definition of $F$ above, making sure that we also rescale $x$ as above.}
We find the metric
\be\label{penr13}
ds^ 2= 2 du dv + dx^2 + d\vec x_3^2 + d\vec x_4^2  + ( F_x  x^2 +
F_3 \vec x_3^2  + F_4 \vec x_4^2 )du^2 \ ,
\ee
where
\ba
&& F_x = -J^2 \left(1- 3 {a^2-1\ov (1-a \cos 2 J u)^2}\right)\ ,
\nonumber\\
&& F_3 = -J^2 \left(1 +  {a^2-1\ov (1-a \cos 2 J u)^2}\right)\ ,
\label{brii}\\
&& F_4= -J^2 \ .
\nonumber
\ea
Note that the metric depends explicitly on light-cone
time $u$. From the discussion around \eqn{general}
and \eqn{einstein} and the form
\eqn{penr11}, we understand
that this background has an $SO(3) \times SO(4)\times U(1)$ symmetry.
The $U(1)$ factor is not manifest in the Brinkman coordinates, in contrast
to the Rosen-like coordinates \eqn{penr11}.\footnote{In general, in
the notation of footnote 2, the $U(1)$ symmetry acts non-trivially
as: $\d x = \sqrt{A}\e $ and $\d v= -\ha x A^\prime/\sqrt{A}  \e$,
where $\e$ is an infinitesimal constant parameter.}
We also find that the Penrose limit taken for the
five-form in \eqn{r2uu1} gives an expression of the form \eqn{general},
namely
\be\label{bono}
\mathcal{F}_5 = J du \wedge (1 + \ast_8)(dx \wedge dx_1 \wedge dx_2
\wedge dx_3) \ .
\ee
It is quite remarkable that the five-form in the particular PP-wave limit that
we have taken does not depend at all on the non-extremality parameter $\m$ and
on the vev parameter $r_0$, but, instead, it retains the form it has for
the maximally supersymmetric PP-wave solution.
This has, as we will see, the important consequence that the fermionic
spectrum in the Green--Schwarz action can be immediately determined.
In addition, in this case, the temperature
effects are washed out when the Penrose limit is taken.
This is also seen by the fact that, in the metric,
$\m$ and $r_0$ always appear combined into the constant $a$
defined in \eqn{fuga}. Therefore, in order to recover
the maximally supersymmetric PP-wave solution
it is not required that both $\m$ and $r_0$ be
set to zero, but simply that $r_0=J \m^2$.

\subsection{Geodesics at $\th=\pi/2$ and zero temperature}

It is also of interest to consider the extremal limit $\mu \to 0$.
In this case the background \eqn{r2uu1} becomes
\ba
&& ds^2  = {r (r^2+r_0^2 \cos^2\th)^{1/2}\ov R^2} \eta_{\m\n} dx^\m dx^\n\
+\ {R^2\ov  r (r^2+r_0^2 \cos^2\th)^{1/2}}
\nonumber\\
&&
\times \left( (r^2+r_0^2\cos^2\th)
\Big({dr^2\ov r^2+r_0^2}+ d\th^2 \Big)
+ (r^2+r_0^2) \sin^2\th d\phi^2
+r^2 \cos^2\th d\Omega_3^2\right)\ ,
\label{ruu1}
\ea
which describes a uniform distribution of D3-branes over a disk
of radius $r_0$, or after we send $r_0 \to -i r_0$,
a uniform distribution of D3-branes over a three-sphere.
Both backgrounds
preserve sixteen supercharges and in the AdS/CFT context they describe
particular points in the Coulomb branch of $\mathcal{N}=4$ SYM in
the regime of large t'Hooft coupling.
Various related studies have been performed in
\cite{Freedman,brandhusfe,brandhusfe2}.

Without extra work we can use the results of the previous
section directly and apply the limit $\mu \to 0$.
The PP-wave metric in Brinkman coordinates have the same form as in
\eqn{penr11} - \eqn{brii}, the only difference being that the definition
for $a$ in \eqn{fuga} has to be replaced by
\be\label{zeroTa}
a = \sqrt{1+4 r_0^2 J^2} ~.
\ee

So far we have assumed that the parameter $r_0^2$ is positive, which
corresponds to a distribution of branes on the two-plane (corresponding to
$\th=\pi/2$ in our parametrization). Hence the geodesic that we have
followed ($\theta=\pi/2$) probes the region near the
brane-distribution. Notice also that the maximum value for $r^2$ is
$(1+a)/(2 J^2)$ which for small $r_0$ becomes $1/J^2$ whereas for large
$r_0$ it becomes $r_0 J$.

If we want to consider the sphere distribution we have to take
$r_0^2 \to -r_0^2$
which implies $a = \sqrt{1-4 r_0^2 J^2}$.
Hence, only for $r_0 J < 1/2 $ the constant $a$ is real and we can keep
our previous formulas for the metrics.
However, for $r_0 J > 1/2 $ the constant $a$ becomes imaginary and the
expressions become unphysical. In particular, there are no real
solutions to the geodesic equation (eq. \eqn{geod} with $\mu=0$) and
the PP-wave metric and the RR five-form flux becomes imaginary.

\subsection{Geodesics at $\th=0$}

The procedure is very similar to the $\th=\pi/2$ case, so we will be brief.
In \eqn{r2uu1} we replace $d\Omega_3^2$ by
\be
d\Omega_3^2 = d\omega^2 + \cos^2 \omega d\psi^2 +
\sin^2 \omega d\tilde\psi^2 ~.
\ee
We take the geodesic located at $\omega = \tilde\psi = 0$ and
the non-trivial angular coordinate along the geodesic is $\alpha = \psi$.
The non-zero coefficients of
the effective three-dimensional metric \eqn{flox} are
\be
G_t = \frac{r^2 (r^2+r_0^2)-\mu^4}{r \sqrt{r^2+r_0^2}} \ ,
\quad G_r = 1/G_t \ ,
\quad G_\psi = 1/\sqrt{1+{r_0^2\ov r^2}} \ .
\ee
The null geodesic equations \eqn{geod}
can easily be solved
\be
r^2(u) = {1 \ov 2 J^2}(1-J^2 r_0^2 - b^2 \cos 2 J u) ~,
\ee
with $b^2 = \sqrt{(1-r_0^2 J^2)^2 + 4 J^4 \mu^4}$.

As before, we make the change of variables \eqn{chava3} and
let
\be
\th = {z \ov R} \ , \qq \omega = \frac{\tilde z}{R} \ ,
\qq  x_{1,2,3} \to x_{1,2,3}/R\ .
\ee
In addition, we define $dz^2 + z^2 d\phi^2 = d\vec x_2^2 $,
so that $z^2 =\vec x_2^2$, and
$d\tilde z^2 + \tilde z^2 d\tilde\psi^2 = d\vec {\tilde x}_2^2 $,
so that $\tilde z^2 =\vec {\tilde x}_2^2$. We also assemble the three
spacelike brane directions into the three-vector $\vec x_3$.
In the Penrose limit $R\to \infty $ the metric takes the form
\be\label{diddy0}
ds^2 = 2 du dv  + A_2(u) d\vec x_2^2 + \tilde A_2(u)  d\vec {\tilde x}_2^2
 + A_3(u) d\vec x_3^2 + A_x (u) dx^2
- C(u,z,\tilde z) du^2 \ ,
\ee
where the numerous functions are given by:
\ba
 && A_x = {1 \ov \Delta} - J^2 r^2 \Delta +
\frac{J^2 \mu^4}{r^2 \Delta} \ ,
\nonumber\\
&& A_3 = r^2 \Delta  \ ,
\nonumber\\
&& A_2 = 1/{\tilde A}_2= \Delta \ ,
\label{diddy}\\
&& C=J^2 \left( \frac{\vec x^2}{\Delta} + \vec{\tilde x}^2 \Delta \right) -
\frac{\mu^4 r_0^2 \vec x^2}{\Delta (r^4 + r^2 r_0^2 - \mu^4)^2} ~,
\nonumber\\
&& \Delta = \sqrt{1+r_0^2/r^2} \nonumber ~.
\ea
We refrain from presenting the metric in Brinkman coordinates,
since the formulas turn out to be quite cumbersome
and not very illuminating. However, we note that, unlike the PP-wave
corresponding to the geodesic at $\th={\pi\ov 2}$, in this case
temperature effects parametrized by the constant $\m$ remain distinct from
those parametrized by the rotational parameter $r_0$, i.e., $\m$ and $r_0$
do not combine into a single constant, as before.

\subsection{Geodesics at $\th=0$ and zero temperature}

The extremal limit is obtained by setting $\mu \to 0$ in all expressions
of the previous subsection which in particular implies
$b\equiv \sqrt{1 - r_0^2 J^2}$. Now,
transforming the PP-wave \eqn{diddy0} and \eqn{diddy} to
Brinkman coordinates we find in the extremal limit
\be
ds^2= 2 du dv + dx^2 + d\vec x_3^2 + d\vec x_2^2  +  d\vec {\tilde x}_2^2
 + ( F_x  x^2 + F_3 \vec x_3^2  + F_2 \vec x_2^2 + \tilde F_2 \vec
{\tilde x}_2^2 )du^2 \ ,
\ee
where
\ba
&& F_x = -J^2 \left(1- {5\ov 4} {b^2-1\ov (1-b^2 \cos^2 J u)^2}
-{1\ov 4} {b^2-1\ov \sin^2 Ju (1-b^2 \cos^2 J u)}\right)\ ,
\nonumber\\
&& F_3 = -J^2 \left[1
+  {b^2-1\ov 4 (1-b^2 \cos^2 J u)^2}\left({b^2-1\ov \sin^2Ju} -b^2+3\right)
\right]\ ,
\label{brii1}\\
&& F_2= -J^2 \left({b^2 \sin^2Ju\ov 1-b^2 \cos^2 Ju} -
{(b^2-1)(4 b^2\cos^4 Ju-2-(b^2+1)\cos^2 Ju) \ov 4
(1-b^2 \cos^2Ju )^2 \sin^2 Ju} \right) \ ,
\nonumber\\
&& \tilde F_2= -J^2 \left({1-b^2 \cos^2 Ju\ov b^2 \sin^2Ju}
+ {(b^2-1)(4 b^2 \cos^4 Ju-2+(1-3 b^2)\cos^2 Ju)\ov
4(1-b^2 \cos^2Ju )^2 \sin^2 Ju }
\right)\ .
\nonumber
\ea

We see that for $r_0 J \leq 1$ the constant $b$ is real and all
expressions make sense, but when $r_0 J > 1$ then the
background becomes complex and is unphysical.
On the other hand if go to the sphere distribution
(see comments after eq. \ref{ruu1}) via
analytic continuation $r_0^2\to -r_0^2$.
All the expressions we remain valid, with
$b=\sqrt{1 + J^2 r_0^2}$, without restrictions on $r_0 J$ .

For the RR five-form field strength in \eqn{r2uu1},
in the PP-limit, we find that
\ba
\mathcal{F}_5  & = &  f(u) du \wedge (1 + *_8)
(dx \wedge d x_1\wedge d x_2\wedge d x_3)
\nonumber\\
& = & {J\ov 2 b} {1-b^2 \cos 2Ju\ov \sin J u \sqrt{1-b^2 \cos^2 Ju}}
 du \wedge (1 + *_8) (dx \wedge d x_1\wedge d x_2\wedge d x_3)\ ,
\ea
which unlike the $\th=\pi/2$ case depends explicitly on $u$.
This can be read off from the equation of motion \eqn{einstein}
which gives the relation
\be
8 f^2(u) =  -(F_x +3 F_3 + 2 F_2 + 2 \tilde F_2) \ .
\label{eqql}
\ee

For the case of the sphere distribution, {\it i.e.} $r_0^2 >0$,
we may consider the limit $b\to \infty$.
The naive limit leads to a well defined expression
for the metric (see \eqn{brii1}), but  $\mathcal{F}_5$ becomes imaginary.
The reason is that $\cos\inv (1/b) \leq J u\leq \pi/2$, which means
that $J u$ is pushed towards the value $J u=\pi/2$.
Therefore, we will consider the following correlated limit
\be
J u \to \pi/2 - u/b\ ,\qq v\to -b J v\ ,
\ee
followed by the limit $b\to \infty$. Then the metric and five-form are well
behaved and the metric maintains its original form, but with
\ba\label{rudy}
&& F_x={6-u^2\ov 4(1-u^2)^2} \ , ~~~
F_3=- {u^2+2\ov 4(1-u^2)^2} \ ,
\nonumber\\
&&F_2= {3(u^2-2)\ov 4(1-u^2)^2} \ , ~~~
\tilde F_2 =   {3 u^2+2\ov 4(1-u^2)^2} \ ,
\ea
where the variable $0\le u\le 1$.
The five-form field strength in this limit becomes
\ba\label{sepp}
\mathcal{F}_5 = & = &  f(u) du \wedge (1 + \ast_8)
(dx \wedge d x_1\wedge d x_2\wedge d x_3)
\nonumber\\
& = & -\ha {1\ov \sqrt{1-u^2}}
 du \wedge (1 + \ast_8) (dx \wedge d x_1\wedge d x_2\wedge d x_3)\ .
\ea
It can be checked directly that \eqn{eqql} is satisfied.


\section{The perturbative string spectrum}

In this section we study the perturbative string spectrum by studying the
corresponding two-dimensional Green--Schwarz action.
The only background fields that couple to the bosonic string modes
are the metric and the NS--NS two-form, where the latter is zero
for our backgrounds. Hence, the bosonic part of the Green--Schwarz action is
just the Polyakov action:
\be\label{poly}
S_B = \frac{1}{4 \pi}
\int d\tau d\sigma g_{\mu\nu} \partial_a x^\mu \partial^a x^\nu ~,
\ee
with $\sigma \sim \sigma + 2 \pi$.

For PP-waves the most natural gauge choice is the light-cone gauge
\be
u= P_+ \tau \ ,
\ee
where, without loss of generality,
a possible additive constant has been set to zero.
If the  metric is of the form \eqn{general} the fluctuations
of the eight transverse modes are governed by
Schr\"odinger equations (see, for instance, \cite{steif})
\be\label{erwin}
-{d^2 x^i_n\ov d\tau^2 } + P_+^2 F_i(P_+ \tau) x^i_n = n^2 x^i_n\ ,
\qq (\rm no\ sum\ over \ i)\ ,
\ee
where we have performed a Fourier transform on the $x^i$
in the world-sheet direction $\sigma$.
Note that this formula is only valid if $F_{ij}$ is diagonal which
is the case for our backgrounds.

The Schr\"odinger equations for the scalar modes with potentials
given by the metric functions $F_i$ in the previous section are in general
hard to solve analytically. However, in many cases it is possible to make
connection with supersymmetric quantum mechanics (SQM) \cite{sqm}.\footnote{
For a comprehensive review on this subject see \cite{cooper}.}
This is a consequence of the relation between the $F_i$ and $A_i$ functions
appearing in the metrics in Brinkman and Rosen coordinates, respectively.
The crucial formula was given in footnote 2 and is nothing but
the expression of the Schr\"odinger potentials $F_i$ in terms of
prepotentials $W_i$
\footnote{This works straightforwardly only if
the metric in Rosen-like coordinates has no
$du^2$ term for the corresponding coordinate.
For instance, this is the case
for the coordinates $x$ and $\vec x_3$ in \eqn{penr11}, but not for
$\vec x_2$ and $\vec{\tilde x}_2$ in \eqn{diddy0}.
Nevertheless, this can be
done for all coordinates with some extra work.}
\be\label{SQM}
F_i = W_i^2 - W_i' \ ,\qq  W_i = -\ha A_i'/A_i   \ ,
\ee
with ground state wave function
\be
\Psi_0 = \sqrt{A_i} ~.
\label{SQMsol}
\ee
Spectra of SQM are positive definite and a zero-mode exists provided
that the norm $\Psi_0$ is finite {\it i.e.}
$|\int A_i| <  \infty$. There exist large classes of exactly
solvable SQM problems \cite{cooper} and in cases where analytic methods
fail, the WKB approximation for SQM does usually better than for generic
quantum mechanical potentials.
Some exactly solvable cases will be discussed later in this section.
We also note that \eqn{SQMsol} represents a solution to the null
geodesic equations corresponding to the metric in \eqn{general}.

In order to find the fermion spectrum we need to extract the
part of the Green--Schwarz
action that is quadratic in fermions.
The relevant formulas have been presented in several papers in the
literature, e.g. \cite{mettse}.
The generalization of the Dirac operator that appears in the
fermion kinetic term is just the differential operator that appears in the
gravitino variation of type-IIB supergravity. Hence, the fermionic part of the
GS action takes the form
\be
S_F = \int d\tau d\sigma i (\eta^{ab} \delta_{IJ} - \epsilon^{ab} \rho_{3IJ})
\partial_a x^\mu \bar{\theta}^I \Gamma_\mu \mathcal{D}_b \theta^J ~,
\ee
where $\eta^{00} = -\eta^{11} = -1$, $\epsilon^{01}=1$ and
$\rho_3 = \diag(1,-1)$. The derivation
$\mathcal{D}_b$ is the pull-back of the differential operator that
appears in the gravitino variation:
\be
\mathcal{D}_a \equiv \partial_a + \partial_a x^\mu
\left(
\frac{1}{4} \omega_{\mu AB} \Gamma^{AB}+
\mathcal{F}_{ABCDE} \Gamma^{ABCDE}
\rho_0 \Gamma_\mu
\right) ~,
\ee
with $\rho_0 = i \sigma_2$.

For our type of backgrounds \eqn{general} this simplifies to
\be
\mathcal{D}_a
\left( \begin{array}{c}
\theta_1 \\ \theta_2 \end{array} \right)
= \partial_a
\left( \begin{array}{c}
\theta_1 \\ \theta_2 \end{array} \right)
+ \partial_a x^+ \left[
\frac{1}{4} \omega_{+MN} \gamma^{MN} \left( \begin{array}{c}
\theta_1 \\ \theta_2 \end{array} \right) +
\frac{f(u)}{2} \gamma_+ (P_1+P_2)
\left( \begin{array}{c} \theta_2 \\ -\theta_1 \end{array} \right)
\right]\ .
\ee
Using this and the light-cone gauge
the fermion equations become
\be\label{suck}
\gamma_+
\left( \begin{array}{c}
(\partial_\tau + \partial_\sigma)\theta_1 \\
(\partial_\tau - \partial_\sigma)\theta_2 \end{array} \right) =
P_+ \frac{f(P_+ \tau)}{2} \gamma_+ (P_1 + P_2)
\left( \begin{array}{c} \theta_2 \\ -\theta_1 \end{array} \right) ~.
\ee
We expand the fermions in Fourier modes in the $\sigma$ direction.
Note also that in addition to the chirality conditions
$\theta_i = \gamma_{11} \theta_i$ the light-cone gauge implies
$\gamma_- \theta_i = 0$.
In the conventions of \cite{metsaev} that means that out of the
32 components of the $\theta_i$ only the first eight components are
non-zero and we have $P_1 \theta_i = P_2 \theta_i$, where $P_1$ and $P_2$
are the projectors defined in \eqn{porp}. This also follows
immediately from the fact that a condition similar to \eqn{pp29} holds
for spinors $\th_i$ as well.
Consequently we can ignore the $\gamma_+$ factors in
\eqn{suck} and arrange the remaining fermionic degrees of freedom
in two 8-component spinors $\tilde\theta_{1,2}$.
In the last step we reshuffle the total of 16 spinor components of
$\tilde\theta_{1,2}$ and redistribute them in two
8-component spinors $\psi_{1,2}$ such that
the differential equation \eqn{suck} is of the form
\be\label{fermi}
\left( \begin{array}{c}
\partial_\tau \psi_1 \\
\partial_\tau \psi_2 \end{array} \right) =
\left( \begin{array}{cc} -i n & -P_+ f \\
P_+ f & i n \end{array} \right)
\left( \begin{array}{c} \psi_1 \\ \psi_2 \end{array} \right) ~.
\ee
If $f$ is constant this equation can easily be solved and gives
harmonic oscillators with frequencies
\be
\omega = \pm \sqrt{n^2 + P_+^2 f^2} ~,
\ee
with each sign having multiplicity eight.
If $f$ depends non-trivially on light-cone time we can eliminate
one of the fermions in terms of the other at the cost of
introducing a second order differential equation:
\be
\psi_1'' - P_+ f'/f \psi_1' + (n^2 + P_+^2 f^2 - i n P_+ f'/f) \psi_1 = 0 \ ,
\ee
where the prime denotes the derivative with respect to the argument.

\subsection{Spectra en d\'etail}

We start with the PP-wave corresponding to the geodesic with $\th=\pi/2$.
The spectrum corresponding to the $\vec x_4$ directions is like in the
maximally supersymmetric case \cite{metsaev}.
The oscillator frequencies are
\be
\omega_4 = \sqrt{n^2+ J^2 P_+^2} \ ,\qq  n=1,2,3, \ldots \ .
\label{aga}
\ee
For the other directions let $z=2 J P_+ \tau$.
Then, using \eqn{brii} and  \eqn{erwin}, we find the two Schr\"odinger
equations (we will collectively denote by $\Psi(z)$ the corresponding
$x^i_n$'s)
\be
-{d^2 \Psi\ov dz^2 } -{1\ov 4}\left(1+{a^2-1\ov (1-a \cos z)^2}\right) \Psi=
{n^2\ov 4 J^2 P_+^2} \Psi \ , \qq n=0,1,\dots
\label{sc1}
\ee
and
\be
-{d^2 \Psi\ov dz^2 } -{1\ov 4}\left(1-3{a^2-1\ov (1-a \cos z)^2}\right) \Psi=
{n^2\ov 4 J^2 P_+^2} \Psi \  ,\qq n=0,1,\dots \ .
\label{sc2}
\ee
For general $a$ the two potentials are not supersymmetric
partners.

For $a>1$, {\it i.e.} the case of a disk, let us define the constant angle
$z_0=\cos\inv 1/a$ which takes values in the interval $z_0\in [0,{\pi\ov 2}]$.
Then the allowed coordinate ranges are, either $z_0\leq z\leq \pi$
or $-z_0\leq z \leq z_0 $.
Near $z=\pm z_0$ the potentials behave as
$-{1\ov 4}{1\ov (z-z_0)^2}$ and ${3\ov 4}{1\ov (z-z_0)^2}$, respectively,
which means
that there is no unitarity problem in these cases, as expected from SQM.
For $z=\pi$ they go to a constant.
When $a<1$, {\it i.e.} the case of a sphere,
the potentials are nowhere singular.

Now let us consider the limit $a \to \infty$
for which the first coordinate range becomes
$\pi/2 \leq z \leq  \pi$.
Shifting the $z$ variable as $z=x+\pi/2$ we
obtain the range $0 \leq x \leq \pi/2$.
Then the two potentials are of the P\"oschl--Teller type I,
the normalizable solutions are given in terms of Jacobi polynomials
and the spectra are quantized with boundary conditions $\Psi(0) =
\Psi(\pi/2) = 0$. For \eqn{sc1} (with $a\to \infty$) we find
\be
\Psi_m = \cos x \, (\sin x)^{1/2} P_m^{(0,\ha)}\left( \cos 2 x \right) \ ,
\qq \left( \frac{n}{4 J P_+} \right)^2 = (m+1) (m+\ha) \ ,
\label{ecep1}
\ee
and for \eqn{sc2} (with $a\to \infty$)
\be
\Psi_m = \cos x \, (\sin x)^{3/2}
P_m^{(1,\ha)}\left( \cos 2 x \right) \ ,\qq
\left( \frac{n}{4 J P_+} \right)^2 = (m+1) (m+\frac{3}{2}) \ ,
\label{ecep2}
\ee
with $m = 0, 1, 2, \ldots$, for both cases. We note that for $a\to \infty$ the
two potentials in \eqn{sc1} and \eqn{sc2} are supersymmetric partners.
However, the corresponding spectra as given above are different. The reason is
that we have imposed the boundary condition of vanishing wavefunctions at the
end points of the interval $x\in [0,\pi/2]$. However, it can be
shown that these boundary conditions are not obeyed by
both sets of
wavefunctions if they are simply related via the usual rules of SQM.
We see that the constant $P_+J$ needs to be quantized in order for the
wavefunctions that obey the appropriate boundary conditions
to be normalizable.
For generic values of $P_+ J$ we should take the wavefunction to be
trivial, i.e. $\Psi=0$. Hence, for generic values of $P_+J$ we have only
the spectra corresponding to $\vec x_4$ with oscillator frequencies given by
\eqn{aga}. When $P_+J$ is quantized according to \eqn{ecep1} or \eqn{ecep2}
we have in addition the excitations of the corresponding coordinates.

For the second coordinate range, in the limit $\a\to \infty $, the results
for the wavefucntions and the spectra are left unchanged.
Indeed, the coordinate range of $z$ is now $-\pi/2  \leq z \leq \pi/2$ and
the eigenfunctions and eigenvalues turn out to be the same as the ones
in \eqn{ecep1} and \eqn{ecep2}.

Another case that is easy to solve occurs
when $a \to 0$, which can be attained only in the sphere
case. As can be seen from \eqn{brii} we obtain the metric
\be
ds^2= 2 du dv + dx^2 + d\vec x_3^2 + d\vec x_4^2
 -J^2( 4 x^2 + {\vec x_4}^2 )du^2 \ .
\ee
Hence,  three of the transverse scalars are massless and the others are
massive with different masses. The corresponding frequencies are
\be
\omega_{(3)} = \pm |n| \,\qq  \omega_{(x)} = \pm \sqrt{n^2+4 P_+^2 J^2} \ ,\qq
\omega_{(4)} = \pm \sqrt{n^2+P_+^2 J^2} \ ,
\ee
with multiplicities, for each sign, three, one and four, respectively.

The fermionic spectrum on the other hand is very simple and
completely independent of any limit taken.
The frequencies, with multiplicity eight are
\be
\omega = \pm \sqrt{n^2 + P_+^2 J^2} ~.
\ee

Finally, we briefly comment on the spectra for the PP-wave
corresponding to the geodesic with $\th=0$.
The Schr\"odinger potentials appearing in the equation of the
scalar fluctuations can be found in \eqn{brii1}.
They are all physical in the sense that they do not violate the
unitarity bound $-{1\ov 4}{1\ov x^2}$.
In the limit $b \to \infty$
the potentials look much more tractable \eqn{rudy}, however,
also with this simplification we were
not able to find the spectra of the fluctuations
explicitly.
The equations for the fermionic modes \eqn{fermi} with $f$
given in \eqn{sepp} turn out to be equally elusive.

\section {Disk distribution in the limit of large radius}

The fact that the metric \eqn{penr13} is well defined in the limit
$a\to \infty$, suggests that there exists a corresponding limit
of the background \eqn{ruu1} before the Penrose limit is taken.
Revealing the precise connection is the purpose of this section.

Let us take the limit $r_0 \to \infty$ of the metric \eqn{ruu1}
which pushes the radius of the disk to infinity.
The resulting geometry is well defined provided that
we also approach $\th=0$ or $\th=\pi/2$. It is straightforward to see
that in the former case one obtains the metric for D3-branes smeared
uniformly
in two transverse directions and the harmonic function behaves as $1/r^2$,
where $r$ is the radius of the four-dimensional transverse subspace.

However, quite surprisingly, in the case that $\th \to \pi/2$
we obtain a distribution of D3-branes that is uniformly
smeared only over the half-plane. Indeed, consider the redefinitions
\be
r_0\to r_0 R\ ,\qq \th={\pi\ov 2} - {z\ov  r_0 R} \ ,
\qq \phi \to {x_6\ov r_0 R^2}\ ,\qq x^\m \to x^\m R\ ,
\label{llp}
\ee
followed by the limit $R\to \infty$. Then the metric takes the form
\be
ds^2 = r (r^2+z^2)^{1/2} \eta_{\m\n} dx^\m dx^\n + {1\ov  r_0^2
r (r^2+z^2)^{1/2}}
\left( (r^2+z^2) (dr^2 + dz^2) + r_0^2 dx_6^2 + r^2 z^2 d\Om_3^2 \right)\ .
\label{neq1}
\ee
The parameter $r_0$ can be absorbed by rescaling
the coordinates appropriately, so that
we can set $r_0=1$.
In order to reveal the structure of the D3-brane distribution that we
have advertised, consider another change of variables
\be
r=\sqrt{r_5-x_5}\ ,\qq z=\sqrt{r_5+x_5}\ .
\ee
Then the metric becomes
\be
ds^2 = H^{-1/2} \eta_{\m\n} dx^\m dx^\n + H^{1/2} \sum_{i=1}^6 dx_i^2\ ,
\label{neq2}
\ee
where we have defined $x_i$, $i=1,2,3,4$ from $dr_4^2 + r_4^2 d\Om_3^2$,
where $r_4^2 = r_5^2 -x_5^2$.
Therefore, naturally $r_5^2 =\sum_{i=1}^5 x_i^2$.
The function $H$ is
\be
H= \ha {1\ov r_5 (r_5-x_5)} \ ,
\ee
and is indeed a harmonic function in the space spanned by $x_1,\dots , x_5$.
The distribution of D3-branes is obviously uniform in the $x_6$-direction.
However, along the $x_5$-direction the distribution extends only along
the positive axis, i.e. for $x_5>0$. This is in perfect agreement with the
singularity of the harmonic function $H$ which is localized
at $x_1=x_2=x_3=x_4=0$ and $x_5>0$.
This can also be confirmed by a direct computation of the
following integral that results when we consider the smearing of the D3-branes
in the half plane
\ba
&&\int_{0}^{\infty} da \int_{-\infty}^\infty db
{1\ov \left[x_1^2 +\cdots + x_4^2 +(x_5-a)^2 + (x_6-b)^2\right]^2}
\nonumber\\
&& \phantom{xxx} = {\pi\ov 2}
\int_{0}^\infty da {1\ov
\left[x_1^2 +\cdots + x_4^2 +(x_5-a)^2\right]^{3/2}}
\\
&& \phantom{xxxxxxx} ={\pi \ov 2}
{1\ov r_5 (r_5-x_5)} \ .
\nonumber
\ea
Note that the limiting metric \eqn{neq1} (or equivalently \eqn{neq2})
exhibits a
$SO(4)\times SO(2)$ symmetry like the original metric \eqn{ruu1}.

We also not that the limit \eqn{llp}
can also be applied to the non-extremal metric \eqn{r2uu1}. However, the
resulting metric is singular and it does not have a regular horizon.

\subsection{PP-wave limit}

Similar to our previous examples consider the PP-wave limit around a
geodesic situated at $z=0$. The corresponding
three dimensional effective metric is of the general form \eqn{flox} with
$\a=x_6$ and non-zero components given by
\be
G_t= r^2 \ ,\qq G_{x_6}=1/r^2 \ ,\qq G_r =1\ .
\ee
After we multiply the metric by an overall $R^2$, followed
by the rescalings $x^\m\to x^\m/R$ and $z\to z/R$ we get, in the limit $R\to
\infty$, a metric of the form \eqn{penr11}, where the functions are
\ba
 && A_x = 1/r^2 - J^2 r^2 = -J {\sin^2 2 J u\ov \cos 2 J u} \ ,
\nonumber\\
&& A_3 = r^2 =  -{1\ov J} \cos 2 J u \ ,
\label{dghjne}
\ea
Passing to the Brinkman coordinates we get a metric of the form \eqn{penr13}
but with
\ba
&& F_x = -J^2 \left(1- {3\ov \cos^2 2Ju}\right)\ ,
\nonumber\\
&& F_3 = -J^2 \left(1 +  {1\ov \cos^2 2Ju}\right)\ ,
\label{briinn}\\
&& F_4= -J^2 \ .
\nonumber
\ea
which is the $a\to \infty$ limit of \eqn{brii}.

\section{Concluding remarks}

In this paper we have constructed new families of PP-wave solutions of
type-IIB string theory that have light-cone (LC)-time dependent five-form
flux and metric, vanishing three-form fluxes and constant dilaton/axion.
In general these backgrounds preserve sixteen supersymmetries and
the world-sheet action has non-trivial interactions due to the
LC time dependence. The latter implies that the string theory,
at least in a generic background, is not exactly soluble.
Furthermore, we studied
fluctuations of the Green--Schwarz string in these backgrounds
and were able to relate the relevant equations
to supersymmetric quantum mechanics problems. Most importantly, this
guarantees that the spectra have no tachyons and in some cases we were
able to work them out exactly.

We have showed that there are special states in the perturbative string
spectrum for particular values of the light-cone parameter $P_+$.
This is somewhat reminiscent of the special discrete states in two-dimensional
string theory that occur for a particular discrete set of
values of the momentum \cite{KlePo}.
We also note that some of our PP-waves, by being simultaneously
non-trivial and solvable, can be used to elucidate an important general
issue, namely, whether a uniform light-cone gauge choice can be made.
In that respect, we note that there are additional string states related to the
folding of strings on themselves, which are not captured by the theory
obtained if a uniform light-cone gauge choice is made
(for original work on this issue, see \cite{Bars}).
It will be very interesting to pursue  further these and related issues.

A generic feature of our PP-waves is that some of the
$F_{ii}$ components of the metric can blow up. It has been argued
\cite{steif} that this leads to singular string propagation and therefore,
naively, one might think that these solutions are unphysical.
However, we believe that this is not the case.
Note, that these solutions are Penrose limits of backgrounds
used in the AdS/CFT correspondence to describe the deformation of
$\mathcal{N}=4$ SYM by turning on expectation values of scalar fields.
On the field theory side this is the simplest deformation one can think
of and is by no means singular. Furthermore, in the IR the supergravity dual
is not a valid description because the curvature blows up. This is a
reflection of the fact that the deep IR is better described by a set of
free abelian gauge theories,
which do not have a ``good'' supergravity dual.
Since the geodesics we used for the
the Penrose limits probe precisely the region in the deep IR,
it was to be expected that the string theory description might
break down at some point
and there should exist a valid field theory description.\footnote{
A breakdown of the string description was also observed in \cite{gps} for
different backgrounds. In the PP-waves studied in \cite{gps} it was
possible to switch to another dual string theory background, whereas
in our case only field theory is a reliable description in the IR.}
This field theory should be a truncation of $\mathcal{N}=4$ SYM
as in \cite{bmn} where the
scalars now obtain vacuum expectation values in the form of continuous
distributions. It would be interesting to investigate this issue further
by performing a direct field theory computation. In particular, we
suspect that the IR is described by a free field theory.

We should also mention that the Penrose limit
of the non-extremal backgrounds washes out the finite
temperature effects responsible for the existence of the horizon in the
original background which is lost after the limit is taken.
This has been observed also in
previous work and is a consequence of the incompatibility of the
existence of a covariantly constant null Killing vector
with the presence of an event horizon \cite{mukund}.

\section*{Acknowledgements}

A.B. would like to thank J.~Gomis and N.~Halmagyi for discussions.
The work of A.B. was supported in part by the
DOE under grant No. DE-FG03-92ER40701.
K.S. would like to thank I. Bars for a usefull correspondence.
He also acknowledges the financial support provided through the European
Community's Human Potential Programme under contracts
HPRN-CT-2000-00122 ``Superstring Theory'' and HPRN-CT-2000-00131
``Quantum Structure of Space-time'', the
support by the Greek State
Scholarships Foundation under the contract IKYDA-2001/22 ``Quantum Fields
and Strings'',
as well as NATO support by a Collaborative Linkage Grant under the
contract PST.CLG.978785 ``Algebraic and Geometrical Aspects of Conformal
Field Theories and Superstrings''.

\end{document}